\documentclass{elsart}
\journal{} 
\usepackage{epsfig}
\usepackage{amssymb}
\begin{document}

% --------------------------------------------------------
% title
% --------------------------------------------------------

\begin{frontmatter}

\title{Characterization of the first true coaxial 18-fold segmented 
n-type prototype detector for the {\sc GERDA}~project}

\author[a]{I.~Abt},
\author[a]{A.~Caldwell}, 
\author[b]{D.~Gutknecht}, 
\author[a]{K.~Kr\"oninger}, 
\author[b]{M. Lampert},
\author[a]{X.~Liu},
\author[a]{B.~Majorovits\corauthref{cor}}\ead{bela@mppmu.mpg.de}, 
\author[b]{D.~Quirion},
\author[a]{F.~Stelzer}
\author[b]{P.~Wendling}
\address[a]{Max-Planck-Institut f\"ur Physik, M\"unchen, Germany}
\address[b]{Canberra--France, Lingolsheim, France}
\corauth[cor]{Max-Planck-Institut f\"ur Physik, F\"ohringer Ring 6, 
              80805 M\"unchen, Germany, 
              Tel.: +49-(0)89-32354-262, FAX: +49-(0)89-32354-528}

\begin{abstract}
The first true coaxial 18--fold segmented n--type HPGe 
prototype detector produced by Canberra--France
for the GERDA neutrinoless double beta-decay project 
was tested both at Canberra--France and
at the Max--Planck--Institut f\"ur Physik in Munich.
The main characteristics of the detector are given and
measurements concerning detector properties are described.
A novel method to establish contacts
between the crystal and a Kapton cable is presented.

\end{abstract}
\begin{keyword}
double beta decay, germanium detectors, segmentation
\PACS 23.40.-s \sep 24.10.Lx \sep 29.40.Gx \sep 29.40.Wk
\end{keyword}

\end{frontmatter}

% --------------------------------------------------------
% body
% --------------------------------------------------------
\section{Introduction}
\label{section:introduction}
The GERmanium Detector Array, GERDA \cite{proposal}, is designed to search 
for neutrinoless double beta (0$\nu\beta\beta$)--decay of~$^{76}$Ge. The 
importance of such a search is emphasized by the evidence of a non-zero 
neutrino mass from flavor oscillations \cite{oscillations} 
and by the recent claim~\cite{hdhk} 
based on data of the Heidelberg-Moscow experiment. 
GERDA will be installed 
in the Hall~A of the Gran Sasso underground Laboratory (LNGS), Italy. 
The experiment is designed to collect an exposure of about 100$~kg \cdot y$ 
quasi background free. This leads to a requirement of a background 
index of better than 10$^{-3}$~counts/($kg \cdot y \cdot keV$) at the 
Q$_{\beta\beta}$--value of 2039~$keV$. 
 The main design feature of GERDA is to use a cryogenic liquid 
(Argon) as the shield against gamma radiation~\cite{heu}, the dominant 
background in earlier experiments~\cite{hmbg}. High purity germanium 
detectors are immersed directly in the cryogenic liquid which also acts 
as the cooling medium. The cryogenic volume is surrounded by a buffer of 
ultra pure water acting as an additional gamma and neutron shield.

Customized segmented detectors are developed, because  
segmentation permits the identification of photon induced background 
events that have energy deposits at multiple locations (multi-site-events)
over a centimeter scale.
They are different from double-beta events 
which predominantely deposit energy at a single location (single-site-events)
with a typical size of less than a few millimeters \cite{seg_mc}.
An 18--fold segmentation was chosen to optimize the identification
of gamma induced background in the Q$_{\beta\beta}$-region. 
The detector and its read-out cables were developed in collaboration 
between the Max--Planck--Institut f\"ur Physik, Munich, and
Canberra--France, Lingolsheim. Canberra--France produced the detector.
It was tested both at Canberra--France and at the MPI.
The results presented here were all obtained operating the detector
in a vacuum test--cryostat.

\section{The Detector}
\label{section:detector} 
\begin{figure}[t]
%\vspace*{-1cm}
%\hspace{-1cm}
\begin{center}
\epsfig{file=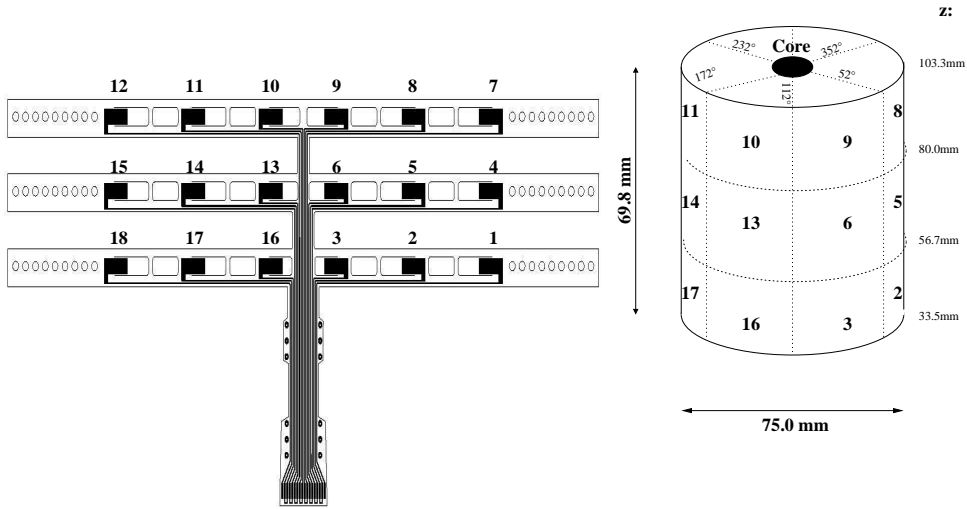,width=13cm}
\caption{Left: Design of the Kapton board  used to 
read out the 
signals of the individual segments. 
Right: Segmentation scheme of the detector including the coordinates 
used .}
\label{fig:segmentation_scheme}
\end{center}
\end{figure}

The detector is n-type with an effective doping
between
$0.7 \times 10^{10} /cm^3 $ and
$1.4 \times 10^{10} /cm^3 $ resulting in a full depletion voltage of
about 2200\,$V$. 
The standard operating voltage was set to 3000\,$V$.
The dimensions 
are 69.8~$mm$ in height and 75.0~$mm$ in diameter.  
The central bore has a diameter of 10.9~$mm$.
The weight is 1.58~$kg$,
the active volume is  302\,$cm^3$ 
The geometry is true coaxial.
The 18--fold segmentation is implemented as
6--fold in $\phi$ and 3--fold in $z$.
The segmentation is created through a 
3-d~implantation process developed by Canberra--France.
The most important detector parameters are listed in 
Tab.~\ref{tab:specifications}.
The numbering of the segments as well as the definition of the
coordinates are given in 
Fig.~\ref{fig:segmentation_scheme}.

\begin{table}[h]
%%%%%\hspace*{4cm}
\begin{center}
\begin{tabular}{|l|r|}
\hline
Parameter & Value \\ \hline
Nominal Operating Voltage & 3000\,$V$\\
Polarity        & positive\\
Crystal Diameter& 75\,$mm$\\
Crystal Length  & 69.8\,$mm$\\
Crystal Mass    & 1632\,$g$\\
Active Volume & 302\,$cm^3$\\
\hline
FWHM at 122 $keV$ & 0.99\,$keV$\\
FWHM at 1332 $keV$ & 1.99\,$keV$\\
Relative Efficiency & 80.4\,\% \\
Peak to Compton Ratio & 75.5\,~ \\
\hline
\end{tabular}
\caption{Delivery specifications of the first 18--fold
segmented true coaxial prototype detector.}
\label{tab:specifications}
\end{center}
\end{table}

\section{The Read-Out Cable with ``Snap-Contacts"}
\label{section:cable} 

The individual segments are read out
using a Kapton printed-circuit-board (PCB).
A schematic is given in Fig.~\ref{fig:segmentation_scheme}.
The contacts between detector segments and board are established with
``snap-contacts".
The Kapton PCB is wrapped around the crystal
with its copper traces on the outside.
It has copper pads which are aligned with the
contact pads of the detector and partially cut out.
They are folded exposing half of their surface
to the crystal pads. 
The Kapton board is tightened 
with a PTFE button that is plugged into
the two appropriate holes of its two ends.
The folded copper pads pressed
onto the detector surface by the PCB form the snap-contacts.

\begin{figure}[h]
\begin{center}
\epsfig{file=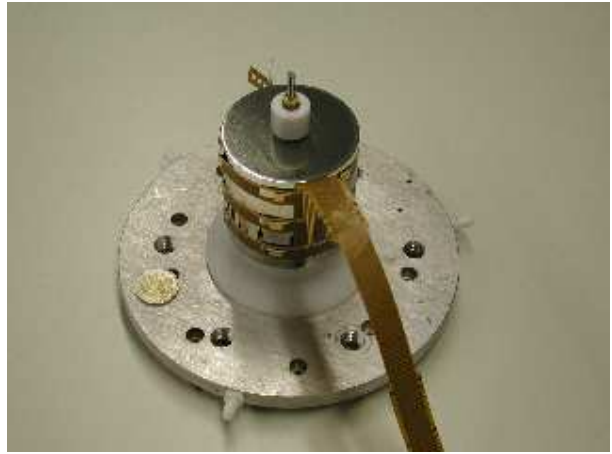,width=8cm}
\caption{Ge crystal with the Kapton board wrapped around.
The 18~segments are contacted with snap--contacts as described in the
text.
}
\label{fig:snaps}
\end{center}
\end{figure}

Kapton and copper thicknesses of 50~$\mu m$ and 35~$\mu m$, respectively, 
were used for this first prototype. 
The tensile strength of the 35 $\mu m$ copper layer 
is essential for the reliability of the contact.
A picture of a bare crystal with a Kapton board attached 
is shown in Fig.~\ref{fig:snaps}.

The snap-contacts establish contact without the introduction of any
other materials. As there is no permanent attachment 
changing of cables is straight forward.

\section{The Test Environment}
\label{section:environment} 

Tests were done both at Canberra--France and at the MPI.
While the tests at Canberra--France focused on the 
basic functionality of the detector 
the measurements at the MPI targeted the
performance with all segments read out simultaneously.

\subsection{The Test Cryostat}
\label{section:cryostat} 

A conventional aluminium vacuum test cryostat was used for all
measurements.
The detector with its Kapton cable and an appropriate high voltage contact
was mounted inside and cooled through a 
copper cooling finger dipped into a conventional 
liquid nitrogen dewar. 

\begin{figure}[]
\begin{center}
\epsfig{file=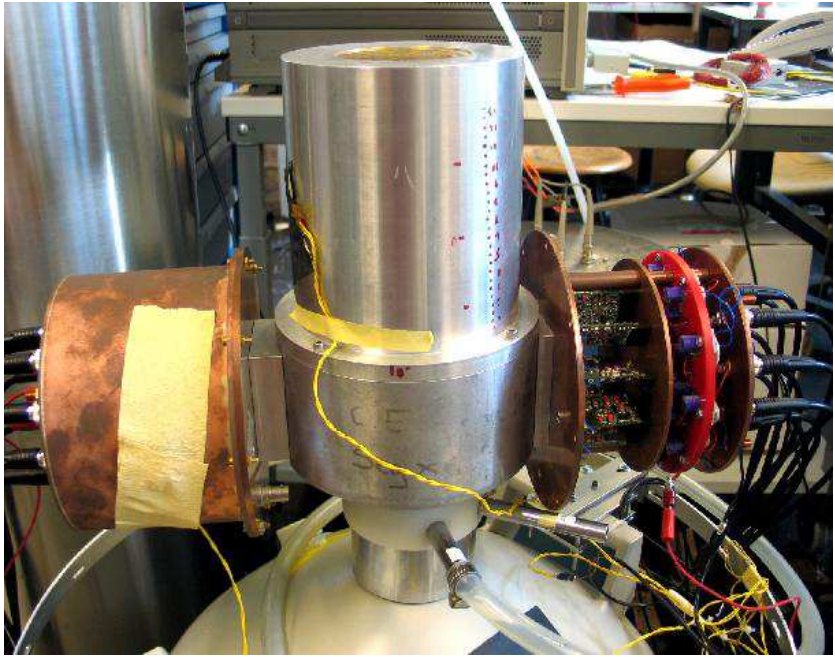,width=11cm}
\caption{
The test cryostat as mounted in Munich on top of 
a 30~$l$ liquid nitrogen dewar. 
The signals left the cryostat through two panels of feed-throughs
and were amplified inside two copper ``ears'' 
housing 19~PSC~823~pre-amplifiers.}
\label{fig:testcryostat}
\end{center}
\end{figure}

The temperature at the top of the cooling finger (the closest
point to the HPGe-crystal itself)
was monitored using a PT100 inside the vacuum cap. 
Between daily refilling the temperature was stable between
-171$^o$\,$C$ and -167$^o$\,$C$.
Spectra within this temperature range were
taken and compared with each other. 
Neither the energy resolutions
nor the calibration parameters changed noticeably.
The vacuum cryostat was pumped to a pressure of $10^{-6}~mbar$
before it was cooled down.
The setup as mounted in Munich is depicted in
Fig.~\ref{fig:testcryostat}.

\subsection{Cabling and Amplification}

\begin{figure}
\begin{center}
\epsfig{file=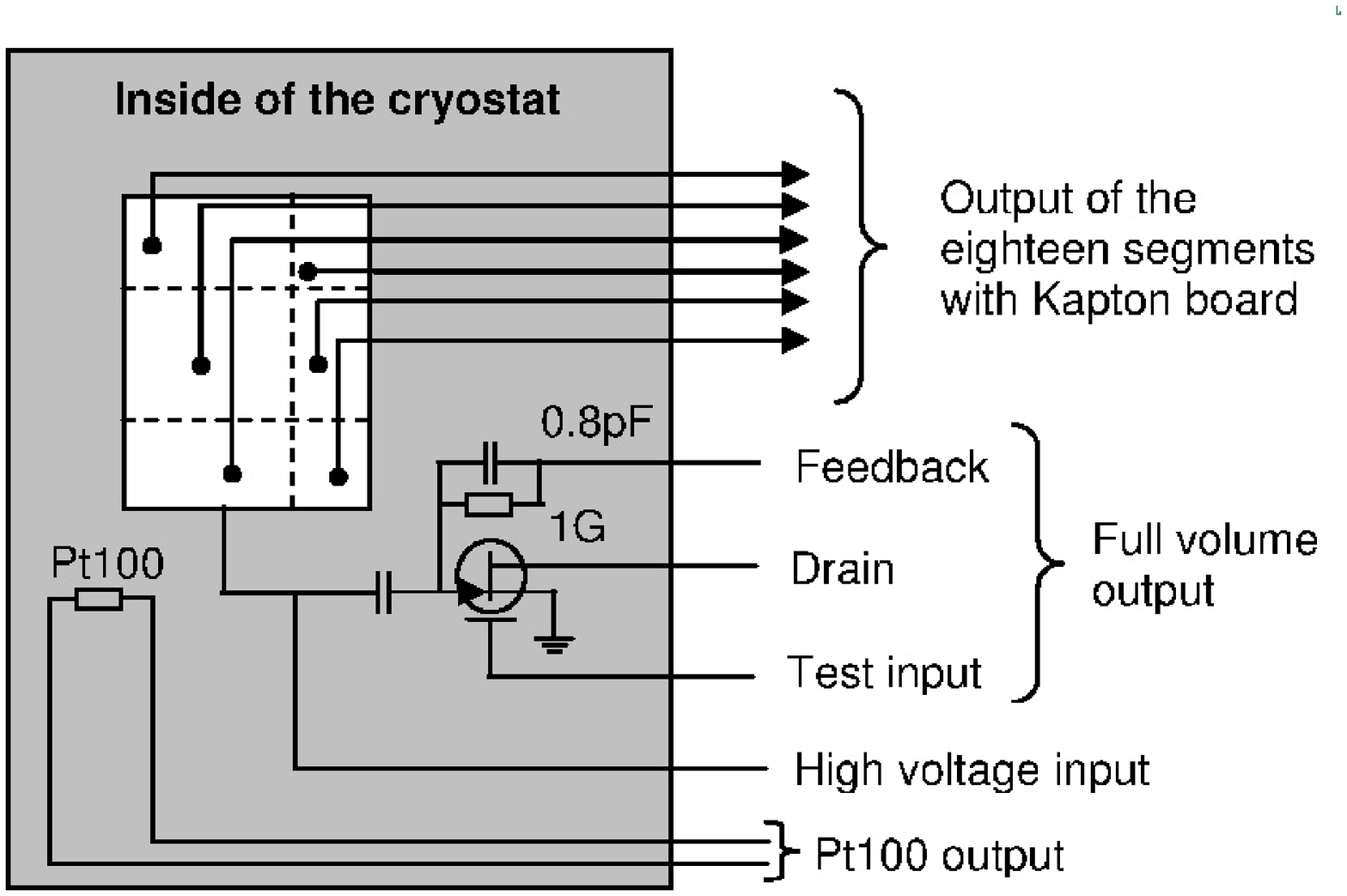,height=5.5cm}
\epsfig{file=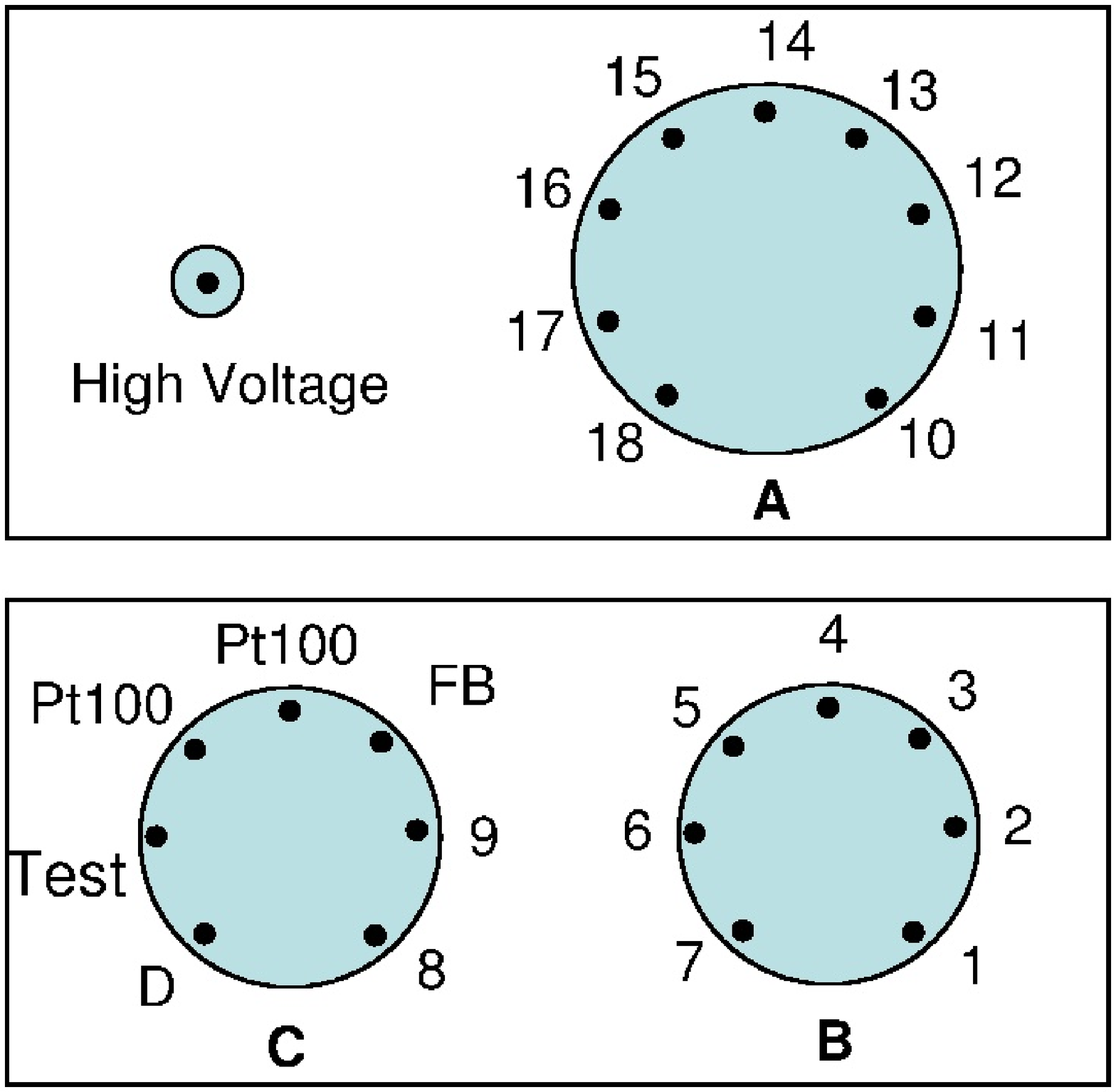,width=5.5cm}
\caption{Left: Schematic drawing of the contacting scheme inside the 
               test cryostat.
         Right: Assignment of the feed-throughs of
               the test cryostat. 
               The high voltage and the signals of segments 10~to~18 
               were serviced on one side.
               The panel on the other side provided
               two feed-throughs for the signals from segments 1~to~9
               as well as
               the drain~[D], feedback~[FB] and test line of the
               FET connected to the core inside the cryostat.
               In addition, the PT100 monitoring 
               the temperature was read out through the
               multi-purpose connector C.
               }
\label{fig:cryo-circuit}
\end{center}
\end{figure}

The test cryostat had two panels of feed-throughs opposite to
each other.
A schematic of the available lines and feed-throughs is
given in Fig.~\ref{fig:cryo-circuit}.
Two 7-channel feed-throughs were located on one side.
Segments 1~to~9 as well as drain and feedback
of the core signal plus the PT100 resistor were connected there.
The signals of segments 10~to~18 were connected to a 
9--channel connector on the panel on the opposite side.
The high voltage input was serviced by 
a separate connector also located on this side.

The signals of the 18~segments were brought out of the cryostat
without any amplification.
Wires were soldered to the Kapton board and to the connectors.
The signal of the core was pre-amplified by an FET inside
the cryostat. All wires inside the cryostat were unshielded.

The signals were always amplified by PSC~823~pre-amplifiers produced
by Canberra.
At Canberra--France single channels were connected to a single
pre-amplifier through a patch-panel.
At the MPI all channels were connected simultaneously.
Figure~\ref{fig:testcryostat} shows the two copper housings
called ``ears'' into which all 19~pre-amplifiers 
were integrated.
The DC coupled room temperature FETs for the signals of the
segments were sitting on the PC~boards of the 18~pre-amplifiers themselves. 
The cold FET for the AC coupled core signal was sitting inside the 
detector cap and was thermally coupled to the cold finger. 

It was of utmost importance to connect the pre-amplifiers to a very well 
defined common ground in order to avoid resonances and noise prohibiting
the operation of the system.
This was achieved through
massive copper plates of 3\,$~mm$ thickness inside the copper ears.
All pre-amplifiers  were directly connected to these plates.
The plates were coupled to the aluminium-cap.
Both the high voltage and the low voltages for the pre-amplifiers
were filtered locally within the ears.

\subsection{Data Aquisition}

At Canberra--France a single channel data aquisition was used.
The signals of the core or of any chosen segment were recorded 
as a histogrammed spectrum.

At the MPI a 75\,MHz XIA Pixie--4 data acquisition system with 5~modules 
with 4~channels each integrated in a National Instruments crate was used.
The pre-amplified signals were filtered and digitized~\cite{pixie}. 
The system has different modes of data storage. 
It can
store energy spectra or for each event time and energy 
information or time and energy 
plus pulse shape information for each channel.

\section{Basic Detector Properties}
\label{section:Measurements}

\subsection{Full Depletion and Operational Voltage}

\begin{figure}[h]
\begin{center}
\epsfig{file=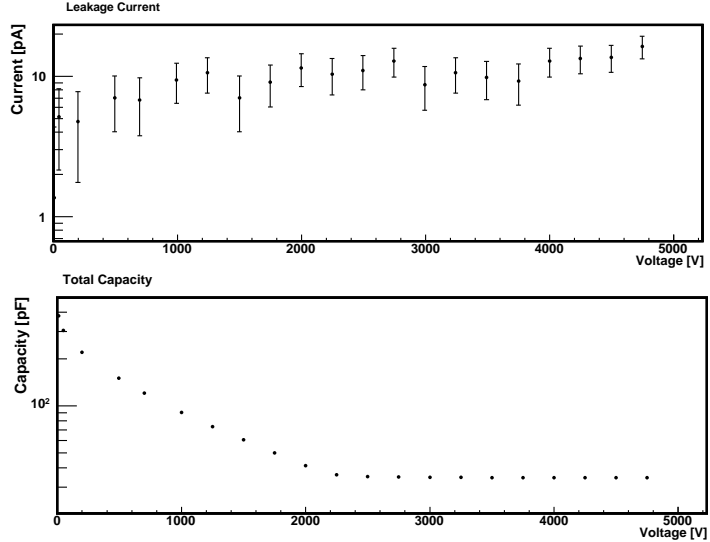,width=11cm}
\caption{Voltage dependence of the leakage current (upper graph)
and  total capacitance (lower graph) of the detector.}
\label{fig:leakage_current}
\end{center}
\end{figure}

The voltage dependence of the
capacitance of the detector and its segments 
as well as of the leakage current were measured
at Canberra--France.
Figure~\ref{fig:leakage_current} visualizes the results.

The leakage current is constant
in the voltage range between
1500\,$V$ and 4000\,$V$. 
Above 4000 $V$ 
the leakage current increases slightly 
with increased voltage.
The detector capacitance decreases 
with increasing voltage
until the value of 35\,$pF$ is reached 
at full depletion around 2200\,$V$.
The standard operational voltage was set to 3000\,$V$.

The 
individual segments 
have capacitances of around 2\,$pF$ at full depletion. 
The values for the detector
as a whole and for all individual segments at 3000~$V$ are listed 
in Tab.~\ref{tab:segments}.

\begin{table}[th!]
\begin{center}
\begin{tabular}{|l|c|c|}
\hline
Segment & Capacitance & Resolution at 1.3\,$MeV$\\
        &  [$pF$]     &   [$keV$]  \\
\hline
\hline
total& 34.88 & 4.07 $\pm$ 0.03 \\
\hline
1&      1.88 & 2.67 $\pm$ 0.05 \\
2&      1.91 & 2.63 $\pm$ 0.06 \\
3&      1.84 & 2.83 $\pm$ 0.05 \\
4&      1.92 & 2.53 $\pm$ 0.03 \\
5&      1.96 & 2.64 $\pm$ 0.04 \\
6&      1.89 & 3.36 $\pm$ 0.05 \\
7&      1.94 & 2.56 $\pm$ 0.02 \\
8&      1.96 & 2.28 $\pm$ 0.02 \\
9&      1.93 & 2.67 $\pm$ 0.02 \\
10&     2.01 & 3.99 $\pm$ 0.03 \\
11&     1.95 & 3.19 $\pm$ 0.03 \\
12&     1.98 & 2.93 $\pm$ 0.02 \\
13&     1.97 & 3.25 $\pm$ 0.01 \\
14&     1.94 & 2.50 $\pm$ 0.03 \\
15&     1.94 & 2.57 $\pm$ 0.03 \\
16&     1.94 & 2.81 $\pm$ 0.10 \\
17&     1.88 & 2.52 $\pm$ 0.05 \\
18&     1.97 & 2.79 $\pm$ 0.07 \\
\hline
\end{tabular}
\caption{Capacitance and energy resolution of each individual 
segment as well as for the total crystal (core in case of resolution). 
Capacitances were measured at Canberra-France, energy resolutions
at the MPI in Munich.}
\label{tab:segments}
\end{center}
\end{table}

\subsection{Resolution}

At Canberra--France the core resolution was determined
from high statistics 
$^{57}$Co and  
$^{60}$Co 
calibration spectra.
The measured FWHM resolutions
were 0.99$~keV$ at 122$~keV$ and 1.99$~keV$ at 1.3$~MeV$, respectively 
(see Tab.~\ref{tab:specifications}).

\begin{figure}[h]
\begin{center}
\epsfig{file=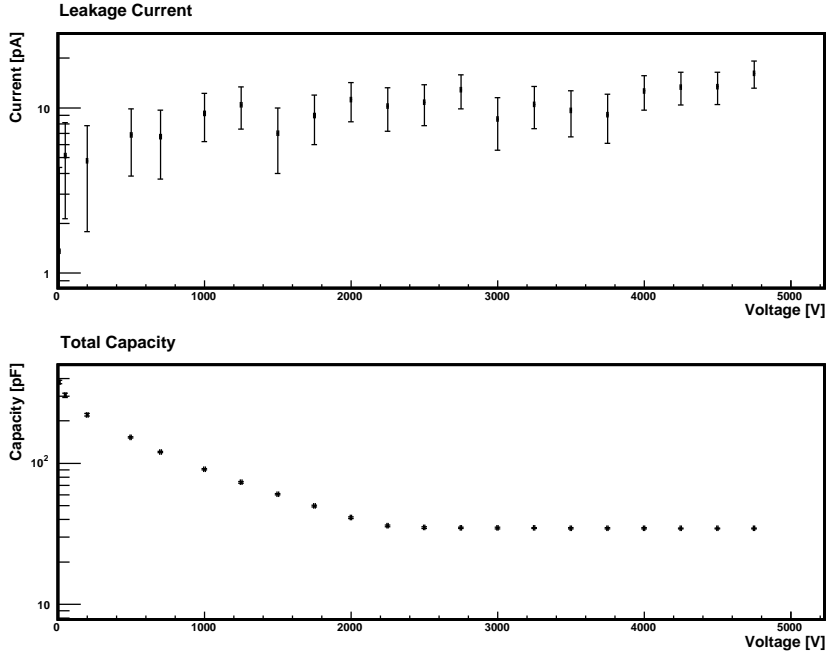,width=13cm}
\caption{Peak area (left panel) and resolution (right
panel) of the 1332.5~$keV$ $^{60}$Co line as a function of the bias
voltage. The energy resolution of the core was 3.1~$keV$ in this measurement
eher not all segments were read out.
In the later measurement referred to 
in Tab.~\ref{tab:segments} it was slightly worse.}
\label{fig:voltage_dep}
\end{center}
\end{figure}

At the MPI the voltage dependence of the core resolution was
studied
with a 60\,$kBq$~$^{60}$Co source located at a distance of 15\,$cm$ to the
side of the test cryostat.
The segments were not read out for this test.
Starting from 2000\,$V$ 
spectra were taken at voltage steps of 100~$V$. 
Each measurement lasted 10~minutes.
The 1.3\,$MeV$ line was used for the analysis. 
For each spectrum a Gaussian plus a constant background was 
fitted and the number of counts within the peak as well as the
FWHM were calculated.
The results are shown in Fig.~\ref{fig:voltage_dep}.
The energy resolution decreases as the count rate increases until
full depletion is reached.

\begin{figure}[h]
\begin{center}
\epsfig{file=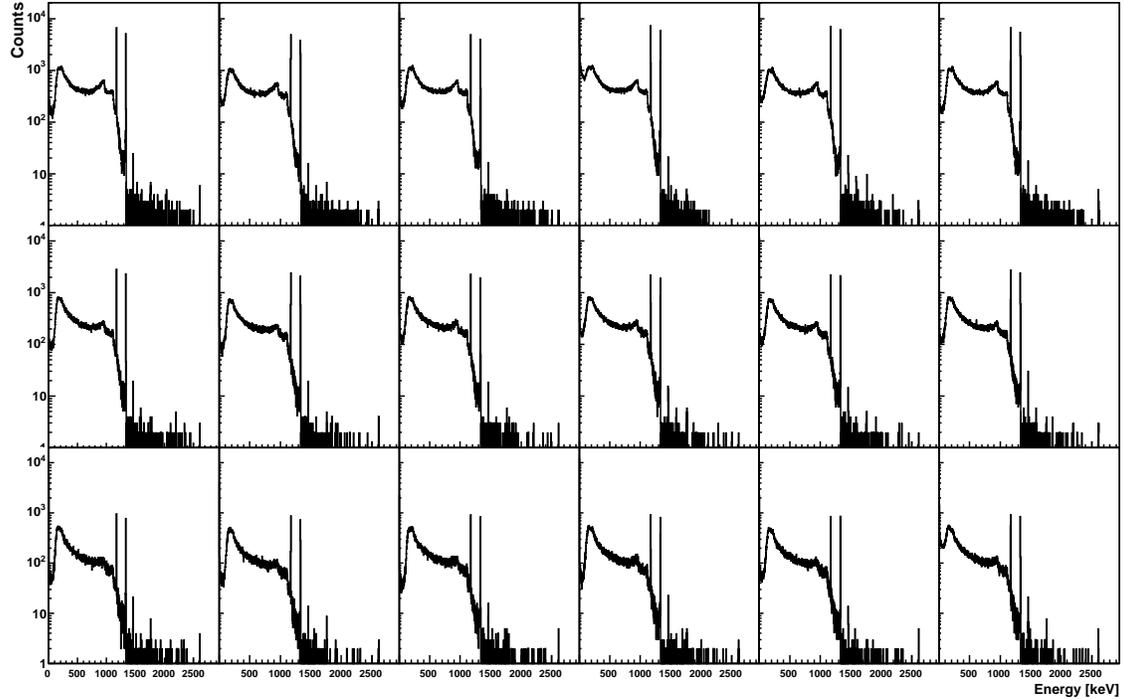,width=15cm}
\caption{$^{60}$Co spectra of all 18 segments when read out
         simultaneously. The arrangement of the graphs is such that
         the spectra of the 6~upper segments are displayed on top,
         the 6~lower at the bottom. The spectra shown are
         corrected for cross-talk (see section \ref{sec:cross-talk}).
         }
\label{fig:spectra}
\end{center}
\end{figure}

The 60\,$kBq$~$^{60}$Co~source was also used to 
determine the resolution of all segments while they were
read out simultaneously.
The source was placed 5~$cm$ above the center of the test cryostat.
The measurement lasted 30~minutes.
The pulse heights of the core and all segments were recorded
for each event. Only events with core energy bigger than
20\,$keV$ were used.
The resulting energy spectra of the individual segments are shown
in Fig.~\ref{fig:spectra}. 
The arrangement of the spectra in Fig.~\ref{fig:spectra} is according
to the physical location of the segments on the detector as depicted in 
Fig.~\ref{fig:segmentation_scheme}.
The upper six spectra correspond to the upper six segments of the crystal
and thus were closest to the calibration source. This is reflected
in the higher statistics in the spectra of these segments.

Fits as described above applied to the 1.3$\,MeV$ 
peak were used to determine the energy resolution of the individual
segments.
The results are listed in Tab.~\ref{tab:segments}.
The segment resolutions range between 2.3\,$keV$ and 4.0\,$keV$.
The core resolution for this measurement was 4.1\,$keV$.
This value of the resolution is worse than the one observed in the
measurement described in a previous paragraph where the segments were
not read out. This is due to interference between the 
experimental surrounding and the electronics channels.

In general the resolutions were not exactly reproducable. 
They varied between 2\,$keV$ and 5\,$keV$.
As mentioned before the performance of the pre-amplifiers was
extremely dependent on the quality of the grounding.
Even the tightness of the screws
closing the ears was seen to change the resolution by the order of
one~keV.
In addition, the setup was exposed
to varying electromagnetic noise 
caused by 
machinery in neighbouring laboratories.
The resolutions were 
definetly not limited by detector properties, but
by the limitations of screening and grounding.

\section{Data Quality}

\subsection{Linearity}

The linearity of the response of the system was tested with
lines distributed between 350\,$keV$ and 2614.53\,$keV$.
The list of lines is given in Tab.~\ref{tab:lines}.
The lines of $^{60}$Co were augmented by several lines of the
background seen in the laboratory.

\begin{table}[th!]
\begin{center}
\begin{tabular}{|r|l l|}
\hline
Energy  [$keV$] & Element & \\
\hline
351.92~$keV$  &  $^{214}$Bi & background       \\ 
609.31~$keV$  &  $^{214}$Bi & background       \\ 
661.66~$keV$  &  $^{137}$Cs & background       \\
1173.24~$keV$ &  $^{60}$Co  & source           \\
1332.5~$keV$, &  $^{60}$Co  & source           \\
2505.74~$keV$ &  $^{60}$Co  & source           \\
1460.81~$keV$ &  $^{40}$K   & background       \\
2614.53~$keV$ &  $^{208}$Tl & background       \\
\hline
\end{tabular}
\caption{The lines used to study the linearity of the read-out system.
        }
\label{tab:lines}
\end{center}
\end{table}

\begin{figure}[t]
\begin{center}
\epsfig{file=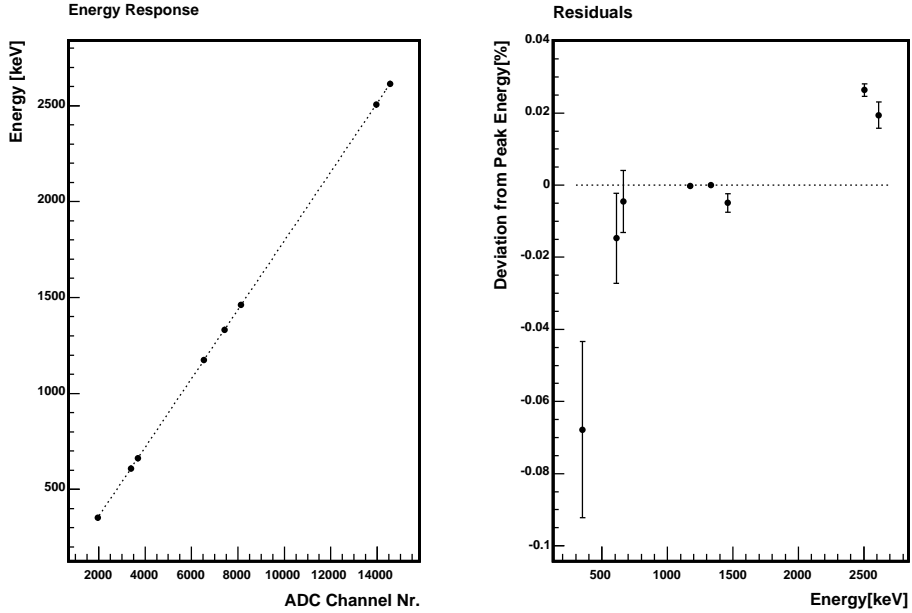,width=13cm}
\caption{Energy calibration of the core signal using peaks from
a $^{60}$Co source and background. The left panel shows the
energy response function. The right panel gives the residuals to
the linear fit function depicted in the left panel.}
\label{fig:calibration_linearity}
\end{center}
\end{figure}

The peaks in the core spectra were fitted with the usual procedure.
The extent of the linearity  can be seen in 
Fig.~\ref{fig:calibration_linearity}. 
The energy response function is plotted in the left panel.
The right panel shows the residuals between the linear fit function
and the data points.

\subsection{Cross-Talk}\label{sec:cross-talk}

\begin{figure}[t]
\begin{center}
\epsfig{file=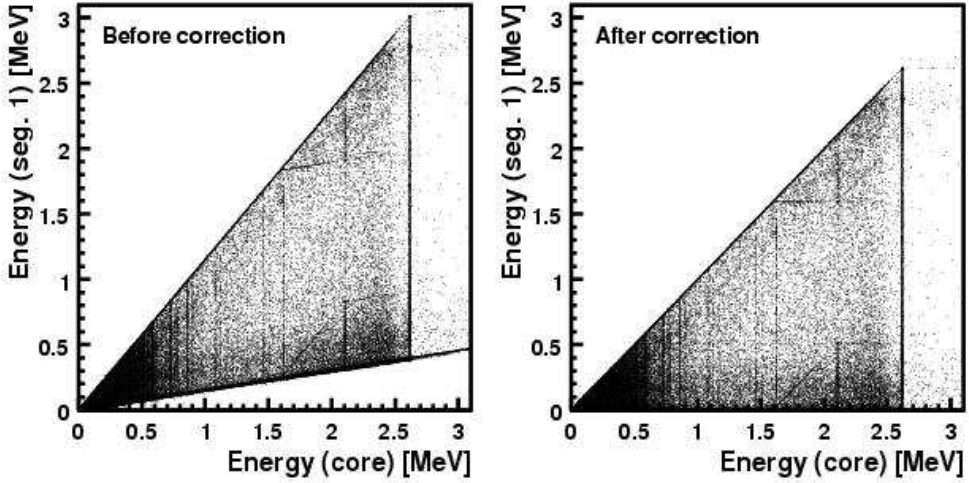,width=13cm}
\caption{Scatter plots of the energy seen by the core [x-axis] vs. 
         the energy seen by segment~1~[y-axis].
         Each dot corresponds to one event. 
         Left panel: Uncorrected data. The cross talk between 
         the signal in the core and in segment~1 is seen
         as the slope of the line below which there are no entries.
         The events with no energy deposited in segment~1 still show
         an effective signal.
         Right panel: Scatter plot of the same data 
         as displayed in left panel, but after 
         cross-talk correction.
        }
\label{fig:scatter}
\end{center}
\end{figure}

Events taken with a $^{228}$Th source were used to study
the cross-talk between read-out channels.
In the left panel of Fig.~\ref{fig:scatter} a scatter plot of 
the energy seen in the core vs. the energy in 
segment~1 is shown.
Each dot corresponds to one event. 
The diagonal which corresponds to the single
segment events with all the energy deposited in segment~1
is clearly visible.
Entries above this diagonal are physically not allowed  
since this would mean that the energy deposited inside a single segment
is larger than in the whole detector.
The core events that did not deposit any energy 
in segment~1 are expected to lie on the x-axis provided 
there is no cross-talk.
In Fig.~\ref{fig:scatter} they are shifted
to a line with a non-vanishing positive slope.
This is due to cross-talk from the core into segment~1.
This is actually expected, since
the FET for the core signal is located inside the aluminium cap. 
Any amplified signal from the core is thus going 
through the same feedthrough as the unamplified signals from segments
1~to~9. This leads to a pick-up in these channels.

The observed cross-talk can easily be corrected for. 
Neglecting second order effects the true energy deposition
in a segment E$_{seg,true}$ can be calculated as

\begin{equation}
E_{seg,true} = E_{seg,meas} - \kappa _{seg} E_{core},
\end{equation}

\noindent
where E$_{seg,meas}$ and E$_{core}$  are the measured
values for the segment and core respectively.
The factor $\kappa _{seg}$ is a correction factor 
to be determined for each individual segment.
Tab.~\ref{tab:x-talk_coeff} lists the coefficients 
for the affected segments.
The cross-talk seen for channels 1~to~9
varies between 1\,$\%$ and 38\,$\%$, being the
strongest for segment~9.
The right plot of Fig.~\ref{fig:scatter} visualizes the
same data as the left plot  after the application of the cross-talk
correction.

\begin{table}[th!]
\begin{center}
\begin{tabular}{|l|r|r|r|r|r|r|r|r|r|}
\hline
Segment & 1 & 2 & 3 & 4 & 5 & 6 & 7 & 8 & 9 \\
\hline
$\kappa_{seg}$  
  & 0.128
  & 0.014  
  & 0.009  
  & 0.033  
  & 0.024  
  & 0.037  
  & 0.014  
  & 0.037  
  & 0.385 \\
\hline
\end{tabular}
\caption{Cross-talk coefficients $\kappa_{seg}$
of segments 1~to~9. Cross talk
from the core signal is seen for the segments that have their
signal feed-through on the same side as the amplified
core signal. No cross-talk is observed for segments 10~to~18.}
\label{tab:x-talk_coeff}
\end{center}
\end{table}

\begin{figure}[t]
\begin{center}
\epsfig{file=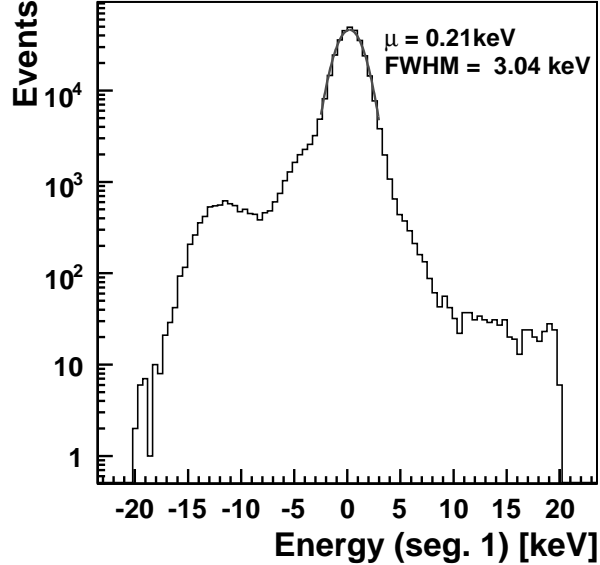,width=8cm}
\caption{Projection of core events
with E~$\geq$~1750~$keV$ onto segment 1 energy after correction.
The same data set as in Fig.\ref{fig:scatter} has been used.
For interpretation see text.}
\label{fig:projection}
\end{center}
\end{figure}

The quality of the correction can be deduced from 
Fig.~\ref{fig:projection}.
The projection of the scatter plot of the right panel of
Fig.~\ref{fig:scatter} onto the y--axis is shown
for events with a measured core energy of more than
1750\,$keV$.
This projection shows the corrected energy in
segment~1 which should be distributed around 0\,$keV$
for the selected events.

A perfect correction would result in only one Gaussian 
peak centered exactly at 0~$keV$. The width would be determined
only by the electronics and the DAQ.
In Fig.~\ref{fig:projection} a Gaussian was fitted for the interval
-3\,$keV$ to +3\,$~keV$. It peaks at  (0.207\,$\pm$\,0.002)\,$keV$ and
results in a FWHM resolution of (3.04\,$\pm$\,0.01)\,$keV$. 
There are second order effects which result in distortions
of the Gaussian. However, they are quite small and do not need to be
corrected for.
Please note the logarithmic scale.

\begin{figure}[t]
\begin{center}
\epsfig{file=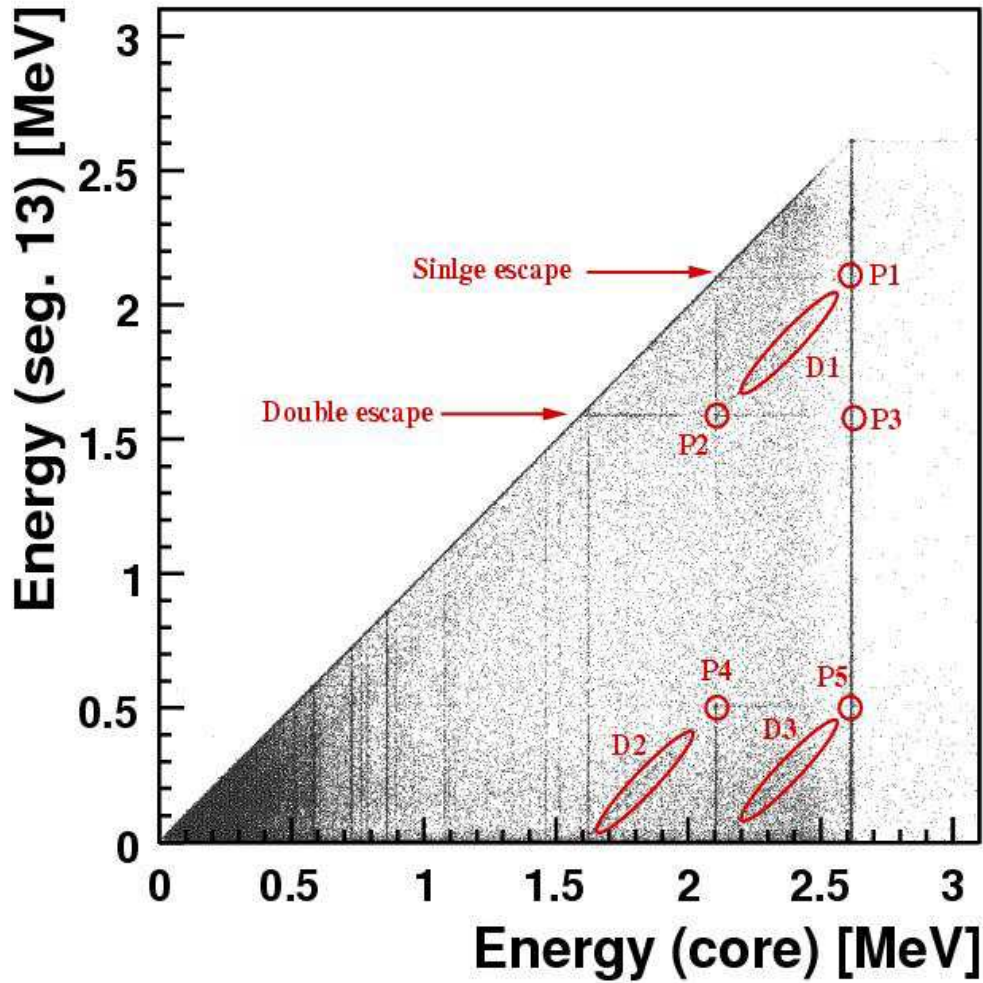,width=13cm}
\caption{Scatter plot of the core energy vs. energy in segment~13. Each
dot corresponds to one event. Cross-talk is not observed
in segment~13.
The marked structures are discussed in sect.~\ref{sec:struct} 
in the text.}
\label{fig:scatter-seg13}
\end{center}
\end{figure}

The segments 10~to~18 are not affected by cross-talk from the core.
Their feed-throughs
are on the other side of the cryostat.
This is demonstarted in Fig.~\ref{fig:scatter-seg13} 
where an uncorrected scatter plot 
is shown for the energy seen by the core
vs. the energy seen in segment~13. 
The cross-talk between segments is insignificant and not taken into
account.

\section{Segment Spectra Correlations}
\label{sec:struct}

Several well understood features appear in the segment~13 scatter plot
depicted in Fig.~\ref{fig:scatter-seg13}. 
The $^{228}$Th~source used has a number of strong gamma lines.
The vertical lines clearly visible
in the plot are caused by events in which
all the energy of a photon is deposited 
within the crystal while
only part of this energy is deposited 
in segment~13.

Photons from the 2615\,$keV$ line  
are responsible for several features marked
in Fig.~\ref{fig:scatter-seg13}.
Photons of this energy are likely to loose their energy via pair production
(see for example~\cite{knoll}). The created
electron will be stopped locally, i.e. with high probability 
within one segment. The positron will also be stopped locally and
annihilate. Two 511~$keV$  gammas are emitted in this process. 
The horizontal lines are due to one ore both 511~$keV$ gammas
escaping the segment but still depositing some energy in the
crystal before escaping it.

The marked areas correspond to the 
following pair-production (pp) scenarios:

\begin{enumerate}
\item[P1:] pp in segment 13. One 511\,$keV$ gamma escapes 
the segment, but is absorbed elsewhere in the crystal.
\item[P2:] pp in segment 13. Both 511\,$keV$ gammas escape the segment. 
One deposits its total energy elsewhere in the 
crystal, the second escapes the crystal.
\item[P3:] pp in segment 13. Both 511\,$keV$ gammas escape the crystal. 
\item[D1:] pp in segment 13. One 511\,$keV$ escapes segment 13 and
deposits its full energy inside the crystal. The other $keV$ gamma
deposits some of its energy in segment 13 and escapes the ctrystal.
\item[P4:] pp inside a segment other than segment~13. One 511 keV gamma
leaves the crystal without depositing any energy. The other 
deposits its total energy in segment~13.
\item[P5:] pp inside a segment other than segment~13. All 2615\,$keV$ are 
deposited in the crystal. One 511~$keV$ gamma deposits its 
total energy in segment~13.
\item[D2:] pp inside a segment other than segment~13. One 511~$keV$
gamma escapes the crystal. The second one deposits 
some of its energy in segment~13 before escaping the crystal.
\item[D3:] pp inside a segment other than segment~13. One annihilation 
gamma deposits some of its energy within segment~13 before 
escaping the crystal. The other gamma deposits all its energy elsewhere
in the crystal.
\end{enumerate}

\section{Detector Characterization}

\subsection{Segment Boundaries}
\label{sec:scan1}

The location of the segment boundaries were determined using
the 122\,$keV$ line of a 
collimated $^{152}$Eu source 
which was moved along  $z$ and $\phi$.
Since the main energy loss mechanism in germanium at the energy
of 122~$keV$
is the photo-effect~\cite{knoll} and the
mean free path is roughly 4~$mm$, the
energy of the photon is deposited predominantly locally.
Segment~14, not affected by cross-talk,
was chosen to be in the center of the test area.
Its center was found to be at $z=68.4$\,$mm$ and $\phi=202^\circ$.
From the center the source was moved in steps of 5\,$mm$ 
in~$z$ and 5$^\circ$ in $\phi$.
The spectra were taken for each measurement  and
the pulse-shapes of all events were recorded.

\begin{figure}[t]
\begin{center}
\epsfig{file=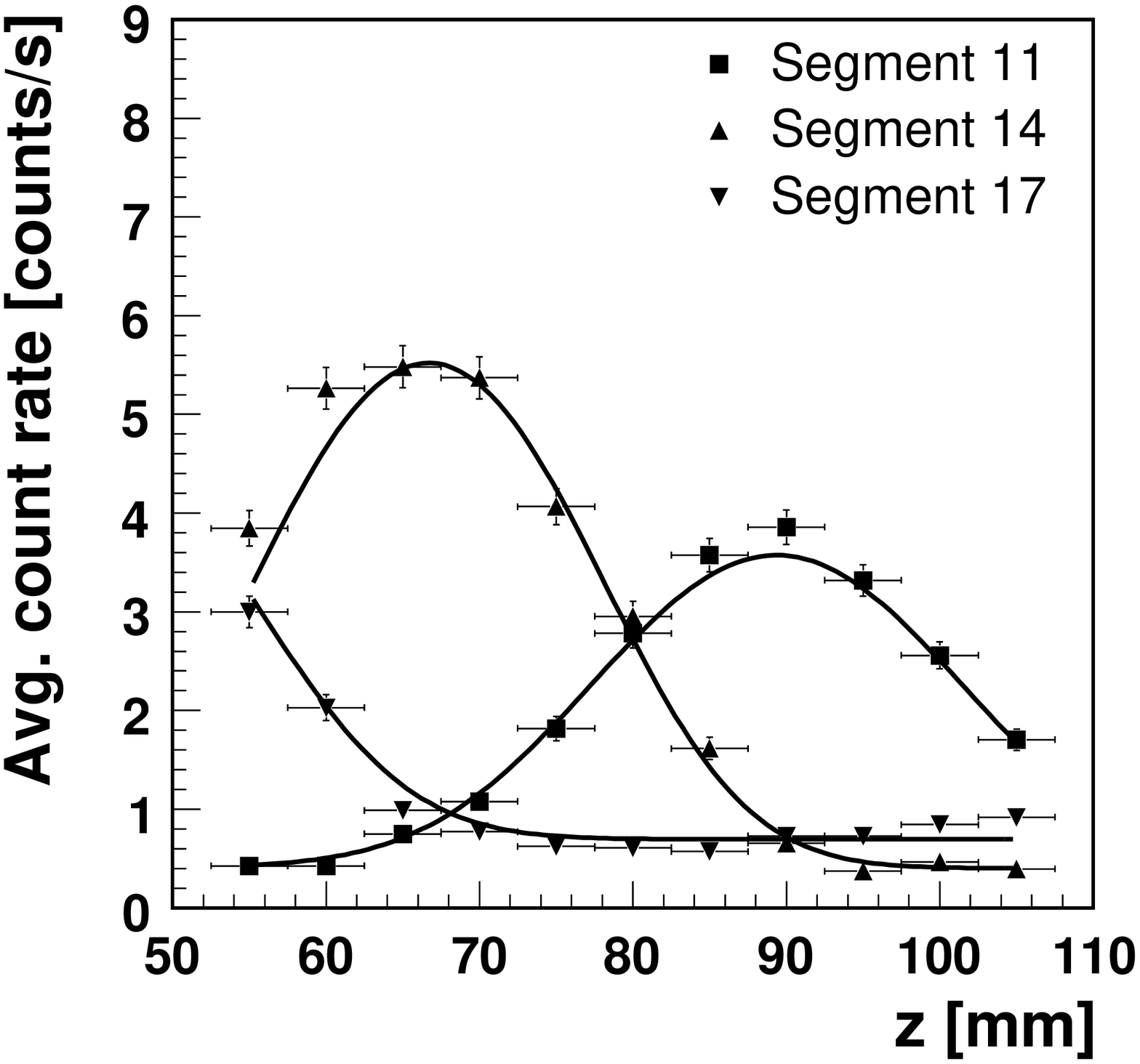,width=6cm}
\epsfig{file=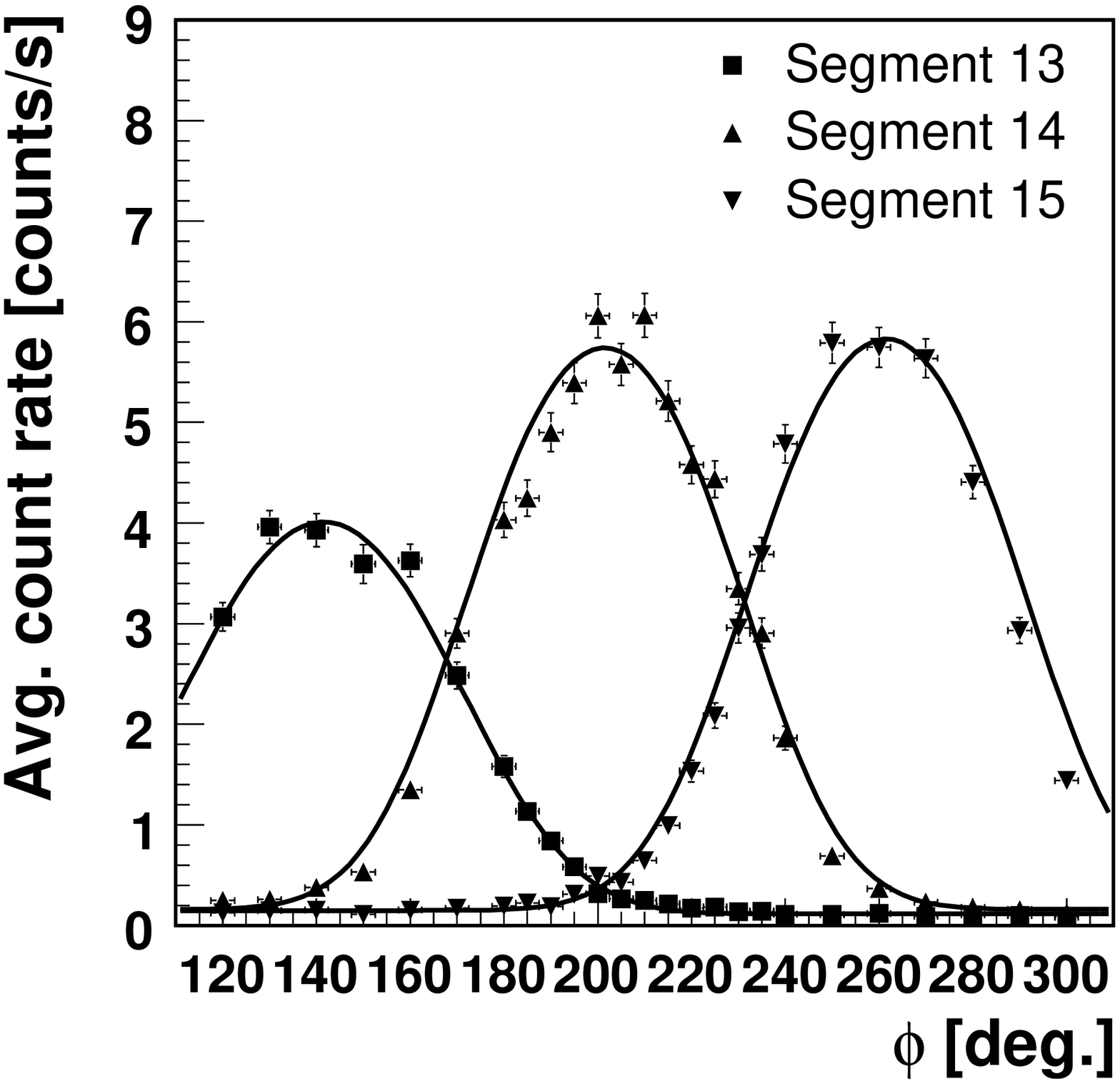,width=6cm}
\caption{Average count rates in dependence of 
         the $z$~[left] and $\phi$~[right] 
         position of the $^{152}$Eu source.}
\label{fig:scanning}
\end{center}
\end{figure}

The average count 
rate in the 122~$keV$ peak was calculated for segment~14 and
its neighbours for each measurement.
The result is shown in Fig.~\ref{fig:scanning}.
The count rates peak whenever the source is placed
above the physical center of a
segment. 
The shape of the count rate distribution can be
described by the convolution of a Gaussian 
representing the width of the beam-spot
and a box function
representing the borders of the segments.
The distances between
maxima were constrained to the known physical distances between
segment centers (23.3\,$mm$ in $z$, 60$^\circ$ in $\phi$). 
In Fig.~\ref{fig:scanning} the results of the fits are shown
for each segment.
As it is determined by the source, 
the size of the beam-spot shoud be constant throughout the measurements.
The widths of the Gaussian functions as fitted are given in
Tab.~\ref{tab:gauss}. The numbers are consistent within errors.
The maxima as seen in Fig.~\ref{fig:scanning} 
can be different for each segment since the
geometry of the aluminium cap as well as the Kapton 
PC-board on the segment do slightly differ between
segments.

\begin{table}[th!]
\begin{center}
\begin{tabular}{|r|c|c|}
\hline
        & Height      & Width      \\
Segment & FWHM        & FWHM       \\
\hline
11 & (5.9 $\pm$ 0.8)~$mm$  & --                   \\
13 &     --                &  (12.3 $\pm$ 0.5)$^o$ \\
14 & (5.6 $\pm$ 0.6)~$mm$  &  (12.0 $\pm$ 0.5)$^o$ \\
15 &     --                &  (13.3 $\pm$ 0.7)$^o$ \\
17 & (6.1 $\pm$ 0.7)~$mm$  & --                   \\
\hline
\end{tabular}
\caption{Size of the beam spot of the $^{152}$Eu source as
         determined from the data used to scan the segment boundaries.
         Within the errors the size stays constant throughout the
         measurements.
        }
\label{tab:gauss}
\end{center}
\end{table}

\subsection{Crystal Axes}

\begin{figure}[t]
\begin{center}
\epsfig{file=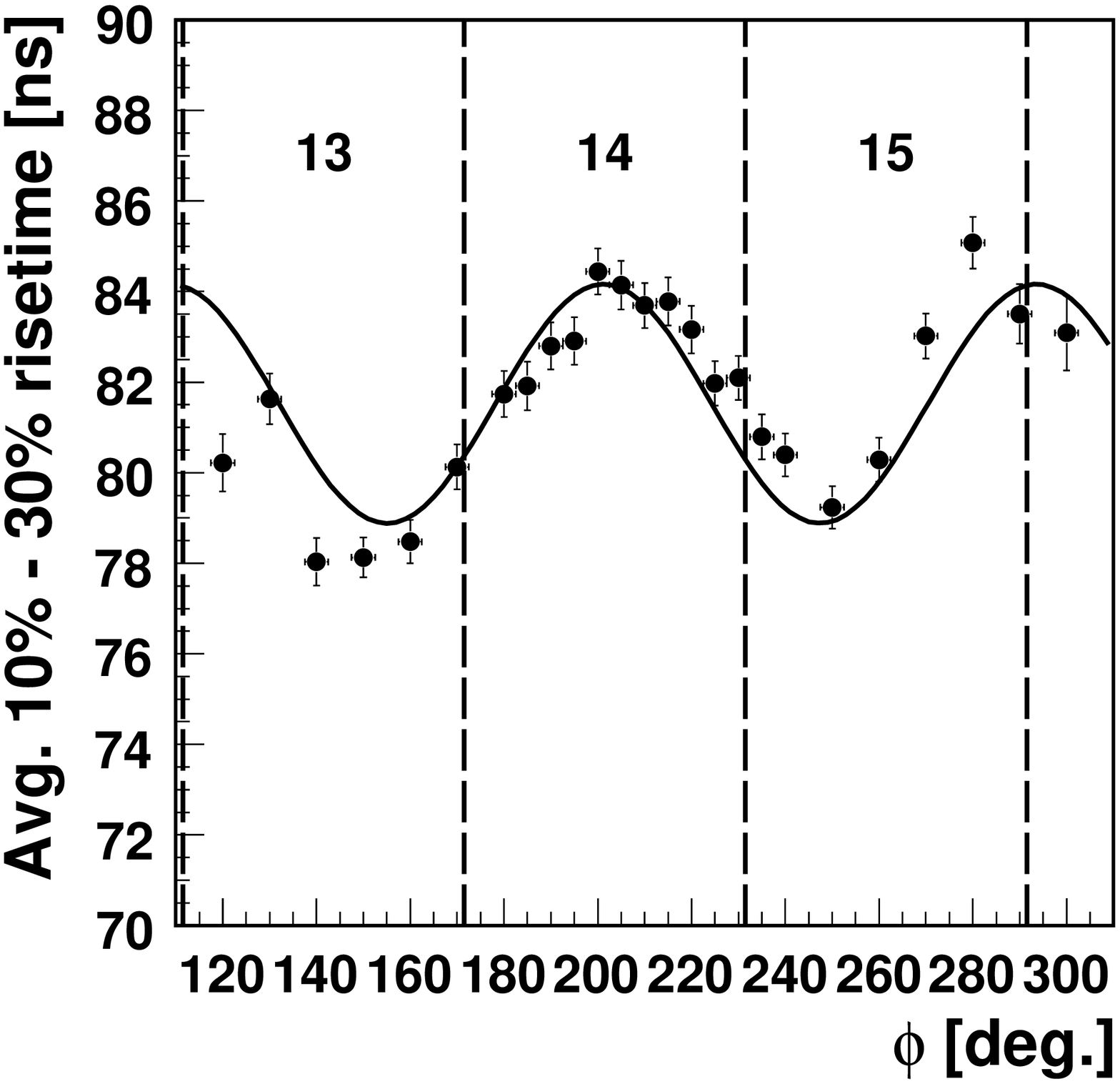,width=6cm}
\epsfig{file=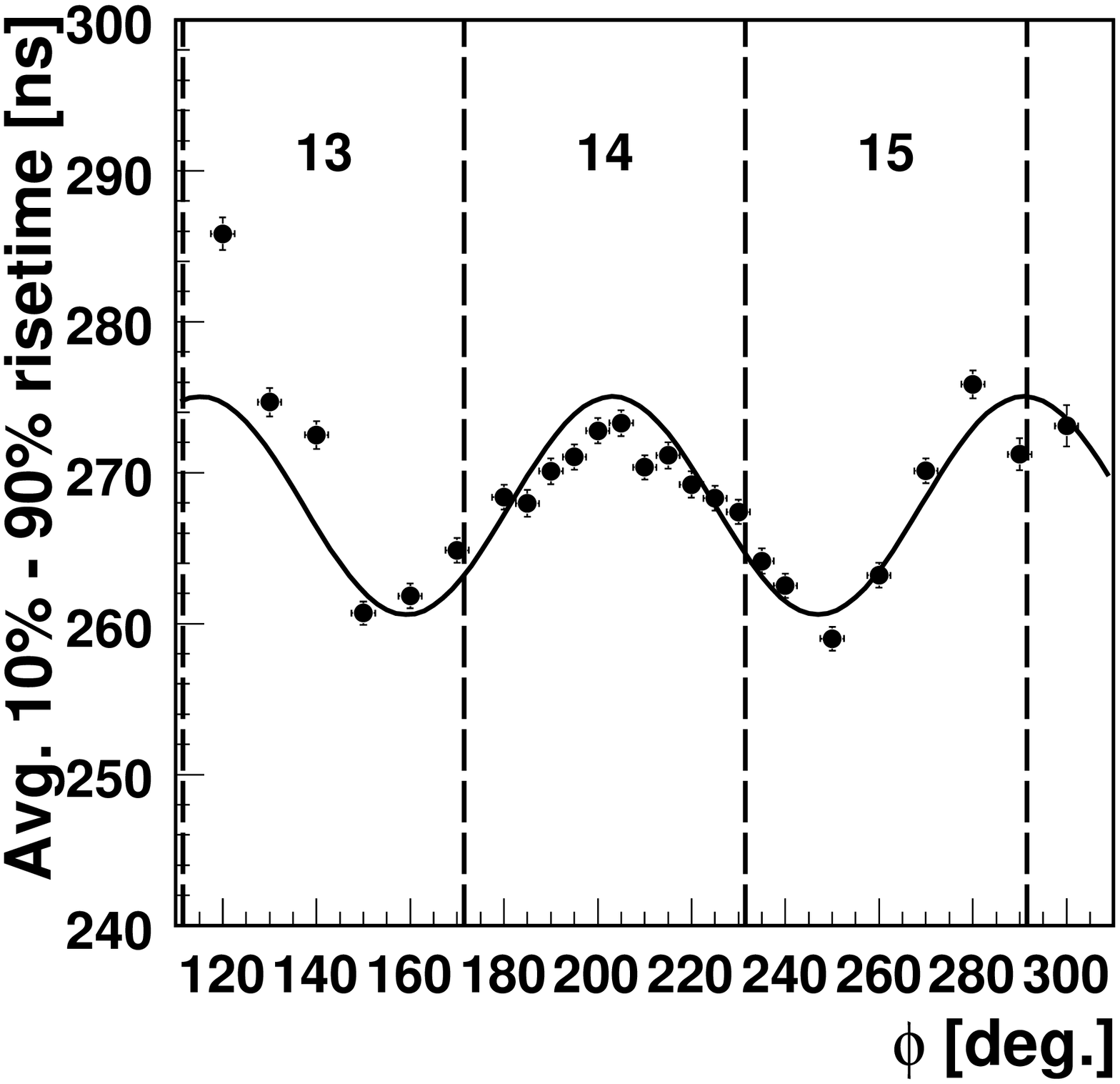,width=6cm}
\caption{Average 10\,$\%$~to~30\,$\%$ (left panel) and 10\,$\%$~to~90\,$\%$ 
(right panel) rise-times as a function of the azimuthal angle $\phi$.
The average rise-times were calculated from pulses in 
the 122~$keV$~line of $^{152}$Eu in which the energy is deposited
very close to the surface.}
\label{fig:rise_time_dep}
\end{center}
\end{figure}

The propagation of electrons and holes in the electric field
of the germanium crystal is influenced by the crystal axes\cite{reik}. 
The charge carriers get deflected and do not follow a simple
radial path.
Thus, the time to collect the charge depends on the angle between
the closest crystal axis and the radial line on which the interaction
takes place.
This is reflected in the rise-times of the pulses recorded.
The pulse-shapes of all events used in the previous section 
were recorded.
The $^{152}$Eu scan along the azimuthal angle $\phi$ was analyzed
with respect to the average 10\,$\%$~to~30\,$\%$ and 10\,$\%$~to~90\,$\%$ 
rise-times of the pulses. The results are shown 
in Fig.~\ref{fig:rise_time_dep}.
The distribution of average rise-times clearly varies with~$\phi$. 
The distance between the extrema is 90$^o$
as expected. The positions of the extrema clearly
establish the orientation of the crystal axes.

\subsection{Radius Dependence of the Rise-Time}

\begin{figure}[t]
\begin{center}
\epsfig{file=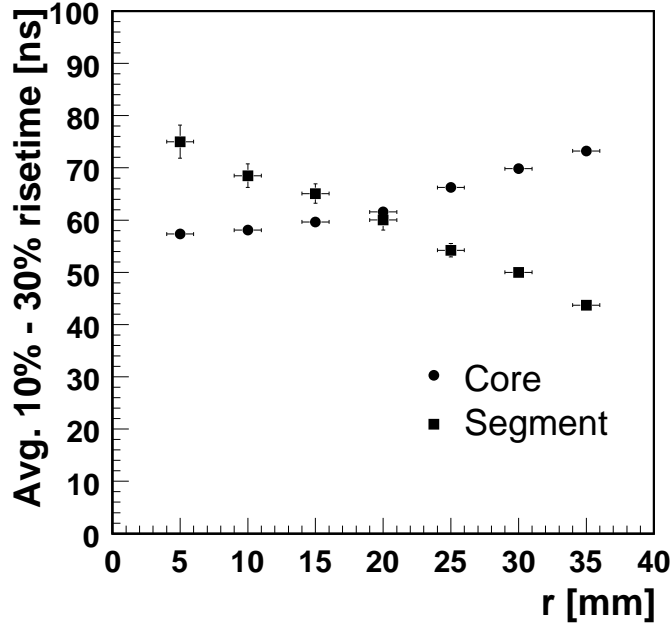,width=9cm}
\caption{Average 10\,$\%$~to~30\,$\%$ rise-time as a function
of the radius of the interaction point for events in the 122~$keV$ peak
of $^{152}$Eu.}
\label{fig:rise_time_r}
\end{center}
\end{figure}

In the previous section interactions close to the outer
surface of the detector were used.
In order to study 
dependence of the rise-time 
on the radius of the interaction point 
the source was placed
above the detector and moved along the center-line
of segment~11 along $r$.  
The end of the collimator with the
$^{152}$Eu source was positioned 5\,$mm$ above the aluminium cryostat. 
Starting at $r=5$\,$mm$ data was taken with a step-size of 5\,$mm$ in $r$.
The pulses in the 122~$keV$
peak were used to calculate the average rise-times for segment~11 and the
core at each 
interaction point. 
Fig.~\ref{fig:rise_time_r} shows a  
clear correlation between rise-time and radius $r$.

The 10\,$\%$~to~30\,$\%$ rise-time was chosen because this separates
between electrons and holes. The average rise-time of the pulses 
from the core increases not strictly linear with the radius. 
At small radii the dependence is very small. 
The dependence of the average rise-time of the pulses from the segment
shows a larger dependence on the radius which, in addition, remains
linear throughout the crystal.
The rise-time of the segment pulses can thus
be used to deduce information about the depth
of interactions.

\subsection{Mirror charge asymmetry}

The electron-hole pairs created in the volume of a given segment
create mirror charges 
in the neighboring segments.
This phenomenon can be
understood and deducted from Ramo's theorem
(see for example \cite{mirror-charges}).
The amplitude of the induced mirror-charge-pulse depends
on the distance between the interaction point and
the boundary between the irradiated segment and the segment under
consideration:
the smaller the distance the larger the mirror charge.
Mirror-charge amplitude asymmetries are defined for the 
neighboring segments opposite to each other:

\begin{equation}
\label{eq:A}
 A_{rl} = log_{10}(A(13)/A(15)) ~~~\mathrm{and}~~~
%\end{equation}
% and 
%\begin{equation}
 A_{tb} = log_{10}(A(11)/A(17)) .
\end{equation}

\noindent
The same data set as in the section~\ref{sec:scan1} was used.
Only pulses in the 122~$keV$ peak were used in order
to select events with mostly only one interaction.

\begin{figure}[t]
\begin{center}
\epsfig{file=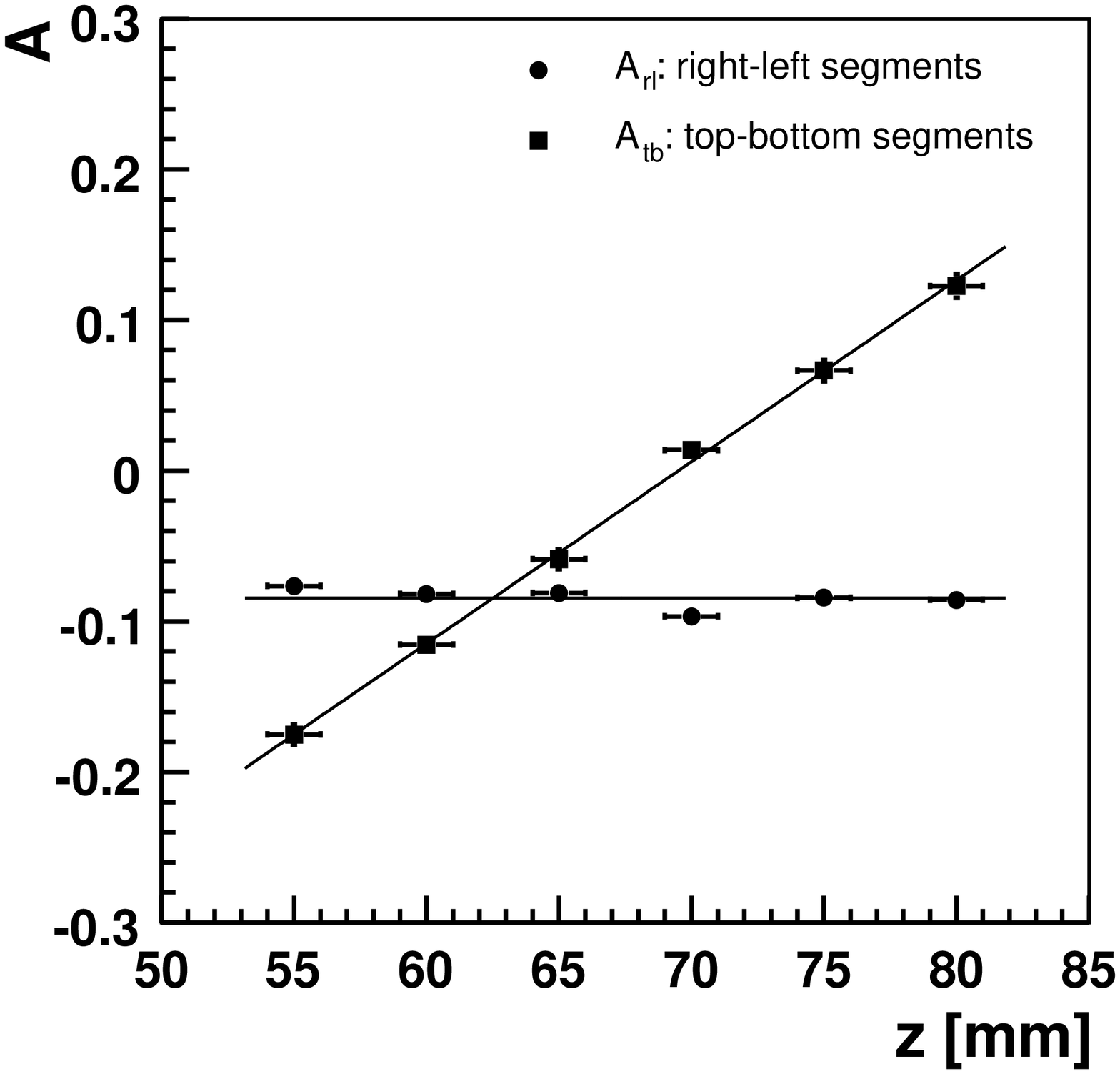,width=6cm}
\epsfig{file=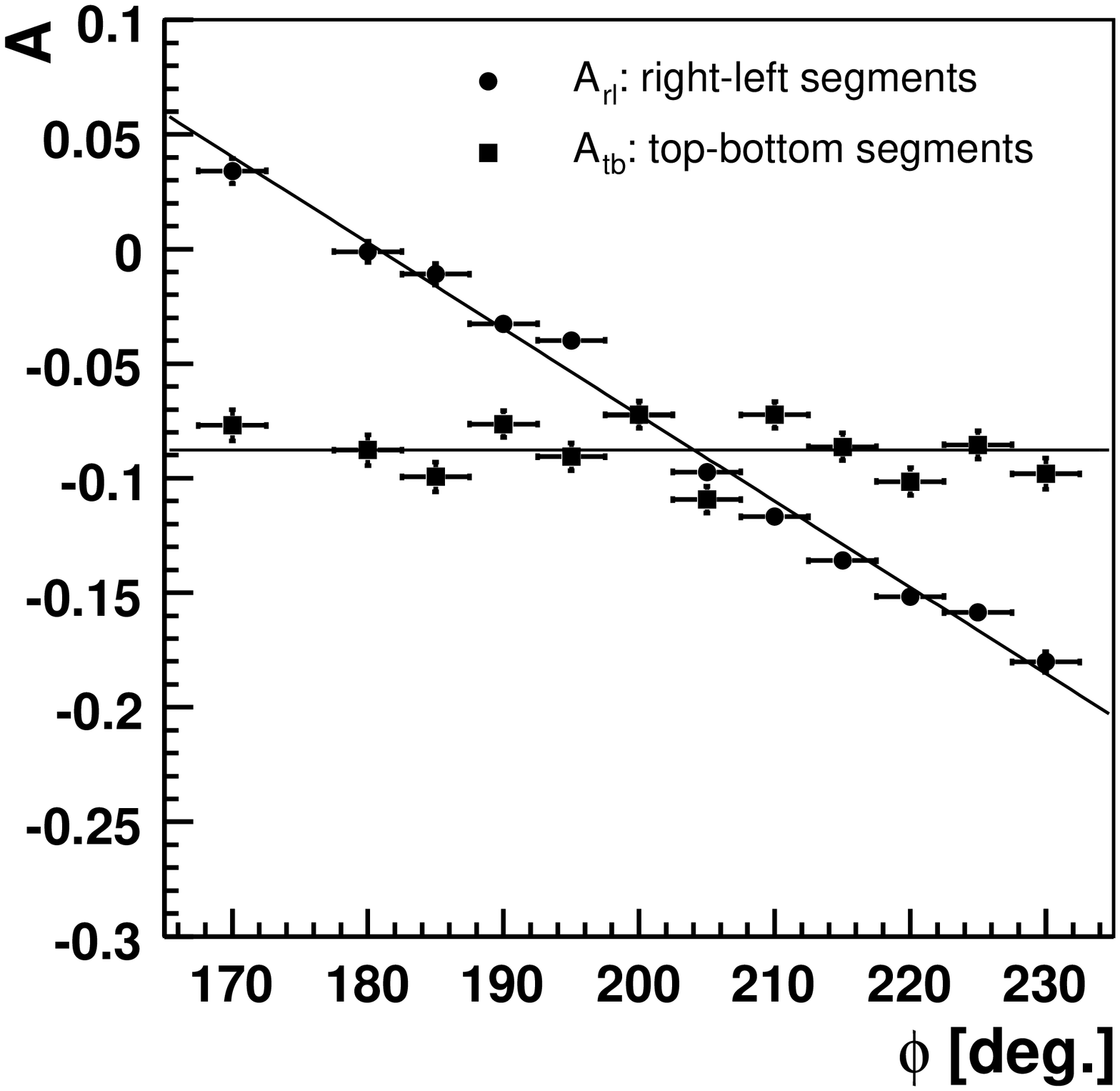,width=6cm}
\caption{Asymmetries between mirror-charge amplitudes of segments
neighboring the irradiated one as a function of~$z$ (left panel) 
and~$\phi$ (right panel). The events in the 122~$keV$ peak
of $^{152}$Eu  were used to calculate the asymmetries A from
equation~\ref{eq:A}.
For the definitions of A and further details see the text.}
\label{fig:mirror-charges}
\end{center}
\end{figure}

In Fig.~\ref{fig:mirror-charges} the mean values of
A$_{rl}$ and A$_{tb}$ are shown for source position as a function of
z and {\it $\phi$}.
In the left panel the measured average 
asymmetries are shown for the z-scan.
The asymmetry of the neighbors left and right from the irradiated
segment (14) is not altered with changing {\it z} since the mean distance
between the segment borders does remain the same for all measurements.
As expected the average asymmetry between neighbors below and above
the irradiated segment does however show a clear dependence on {\it z}.
The difference of the mirror charge amplitudes on
segments 11 and 17 is the largest for the
measurements closest to their boundary.
As shown in the right panel of Fig.~\ref{fig:mirror-charges}
the predicted behavior is also reproduced from the {\it $\phi$}-scan.

\subsection{Position Resolution}

If the mirror-charge amplitude asymmetry distributions are known,
the  $z$- and $\phi$-coordinates of
the interaction point can be obtained for single-site events.
The precision obtained is 6\,$mm$ in~$z$ and about 13$^\circ$ in~$\phi$.
It is dominated by the beam spot size.

Additionally, the radius can be obtained from the radius dependence
of the rise time as shown in Fig.~\ref{fig:rise_time_dep}. The
resolutions is about 5\,$mm$.

\section{Photon Identification}

The ability to identify photon induced events was
studied in detail with several sources.
A detailed analysis is presented in~\cite{gerda_segmentierung}.

\begin{figure}
\begin{center}
\epsfig{file=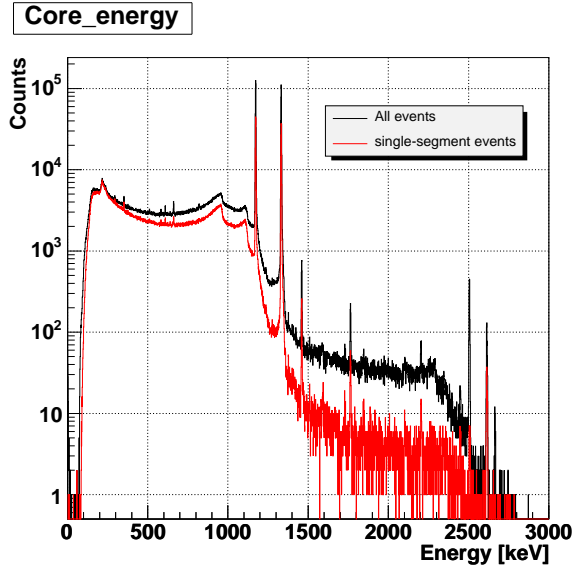,width=8cm}
\vspace{1cm}
\caption{Core spectrum of a $^{60}$Co source placed 5\,cm
above the crystal. The black
histogram shows the total core spectrum, the red one the spectrum of single
segment events. The reduction of the Compton-background
at 2\,$MeV$ is roughly a factor of 10.}
\label{fig:sum_spec}
\end{center}
\end{figure}

Here, only a core spectrum taken with a $^{60}$Co~source 5\,$cm$
above the crystal is shown. Also shown in
Fig.~\ref{fig:sum_spec} is the 
reduction of the spectrum, if it is required that only one
segment has an energy deposition above 20\,$keV$.
At around 2\,MeV about 90\,\% of the events are rejected.
The summation  peak of
the 1173.2\,$keV$ and 1332.5\,$keV~^{60}$Co~lines at 2505.7\,$keV$ is 
almost completely suppressed
in the anti-coincidence spectrum. This is due to the fact that
it is very unlikely for the two photons
to deposit their full energy in the same segment.

\section{Conclusion}

The first 18--fold segmented true coaxial n-type HPGe detector
developed and produced by
Canberra-France, Lingolsheim, 
worked according to specifications within
a conventional test cryostat. 

A novel low-mass contacting scheme was developed and tested.
The copper pads of
a Kapton PC-board were shaped into snap-contacts
which contact the crystal pads without any additional material.
The contacts performed reliably during the tests.

The energy resolutions (FWHM) between 2\,$keV$ and 4\,$keV$
for core and segments were limited by the grounding and
screening of the test electronics.
Crystal axes were determined as well as the radial dependence
of the rise-time of the pulses.
The reconstruction of the position of a single interaction was
investigated as well as the identification of photons with
the help of segment anti-coincidences.
The results confirmed the expectations.

%--------------------------------------------------------
% bibliography
%--------------------------------------------------------

%--------------------------------------------------------
\end{document}